\documentclass[%
 aps,
 amsmath,amssymb,
 reprint,%
]{revtex4-2}

\usepackage{graphicx}
\usepackage{dcolumn}
\usepackage{bm}

\usepackage[utf8]{inputenc}
\usepackage[T1]{fontenc}
\usepackage{mathptmx}
\usepackage{hyperref}
\usepackage[squaren,Gray]{SIunits}
\usepackage{comment} 
\usepackage{url}
\usepackage[normalem]{ulem}
 \usepackage{xcolor} 

\begin{document}

\preprint{AIP/123-QED}

\title[\textit{In situ} sub-50 attosecond active stabilization of the delay between infrared and extreme ultraviolet light pulses]{\textit{In situ} sub-50 attosecond active stabilization of the delay between infrared and extreme ultraviolet light pulses}

\author{Martin Luttmann}
    \email{martin.luttmann@cea.fr}
    \affiliation{Universit\'e Paris-Saclay, CEA, CNRS, LIDYL, 91191 Gif-sur-Yvette, France}


\author{David Bresteau}
    \affiliation{Universit\'e Paris-Saclay, CEA, CNRS, LIDYL, 91191 Gif-sur-Yvette, France}
\author{Jean-Fran\c{c}ois Hergott}
    \affiliation{Universit\'e Paris-Saclay, CEA, CNRS, LIDYL, 91191 Gif-sur-Yvette, France}
\author{Olivier Tcherbakoff}
    \affiliation{Universit\'e Paris-Saclay, CEA, CNRS, LIDYL, 91191 Gif-sur-Yvette, France}
\author{Thierry Ruchon}
    \email{thierry.ruchon@cea.fr}
    \affiliation{Universit\'e Paris-Saclay, CEA, CNRS, LIDYL, 91191 Gif-sur-Yvette, France}

\date{\today}

\begin{abstract}
The blooming of attosecond science (1 as = $10^{-18}$ s) has raised the need to exquisitely control the delay between two ultrashort light pulses, one of them being intense and in the visible spectral range, while the second is weak and in the Extreme Ultra-Violet spectral range. Here we introduce a robust technique, named LIZARD (Laser-dressed IoniZation for the Adjustment of the pump-pRobe Delay), allowing an active stabilization of this pump-probe delay. The originality of the method lies in an error signal calculated from a two-photon photoelectron signal obtained by photoionizing a gas target in an electronic spectrometer with the two superimposed beams. The modulation of sidebands in phase quadrature allows us to perform an \textit{in situ} measurement of the pump-probe phase, and to compensate for fluctuations with an uniform noise sensitivity over a large range of delays. Despite an interferometer length of several meters, we achieved a long term stability of 28 as RMS over hours. This method could be applied to the stabilization of other types of two-color interferometers, provided that one of the propagating beams is capable of photoionizing a target.

\end{abstract}

\maketitle

\section{\label{intro}Introduction}
 Time resolved experiments targeting attosecond precision \cite{Calegari2016, Ueda2019} are commonly based on pump-probe schemes, requiring a stability of the two optical paths better than a few tens of nanometers. For instance, a stability of 3 nm corresponds to a delay resolution of 10 as. These requirements would not be too challenging to fulfill for standard optical interferometers. In attosecond science experiments, the situation is much less favorable for several reasons. In one of the arms of the interferometer, a laser delivers femtosecond light pulses in the Visible-Infrared spectral range (Vis-IR), which are focused in a gas target to undergo a highly non linear process, High Harmonic Generation (HHG). They are thus upconverted to the  Extreme Ultra-Violet (XUV) range \cite{Ferray_1988, McPherson87} and, upon spectral filtering, form attosecond light pulses in the temporal domain \cite{14}. The remaining IR is filtered out downstream the focus, at a sufficient distance to avoid any damage on the filters and optics, usually tens of centimeters to meters away \cite{Ye2020}. In addition, in order to allow the XUV beam propagation, the interferometers are under secondary vacuum. It is the combination of XUV wavelength, highly non-linear stage, meter-size interferometer, and vacuum environment that makes the nanometer-grade stability a challenging  requirement. On top of that, a strong demand is emerging for performing experiments that require long accumulation times (over tens of hours), for instance  coincidence spectroscopy \cite{ColtrimArpes1, Gallmann2017, JJ}, or photoemission from surfaces \cite{Dendzik2020}. In order to meet those needs, two main experimental strategies are combined: increasing the repetition rate of the laser to reduce the acquisition time, and improving, passively or actively, the mechanical stability of the interferometer.

An excellent control of the environment, including temperature, humidity and vibrations, but also the mechanical decoupling of the optical elements from the pumping system, is usually a prerequisite of this second strategy. In addition, the length of the interferometer and the number of optical elements  should be reduced as much as possible. This trend culminated in some original experiments \cite{30weber, 10}, in which the two beams were recombined \emph{before} the HHG setup. Although allowing attosecond pump-probe experiments in fairly difficult conditions, it reduces the flexibility of the beamline and leads to a modulation of the attosecond pulses intensity, which may be difficult to cope with. 
However those efforts are usually not sufficient. Even under such controlled environment, the thermal expansion of the mirrors due to residual absorption of laser light still causes a dramatic drift of the interferometer arms lengths. This is especially true for high repetition rate lasers, asking for active stabilization systems.

Active stabilization systems using feedback loops have been implemented in a series of configurations, all requiring significant modifications of the main setup. Indeed, since the intense IR laser beam generally has to be filtered out from the generation arm after HHG (with e.g. metallic foils), it cannot be recombined to form a standard interference pattern at the output of the interferometer. Any copropagating beam of comparable wavelength cannot either follow the same full path as the XUV beam. Therefore most existing active stabilization techniques rely on an auxiliary continuous wave (cw) laser propagating on a shifted path along the XUV and IR beams in the interferometer \cite{2,3,4,5,6,7}. It yields interference fringes which are  used to perform feedback on the delay by displacing mirrors. Although Vaughan \textit{et al.} \cite{8} developed a different technique allowing a coaxial propagation, with the outer part of the cw beam passing around the filter, it does not result in a simplification of the setup. Indeed, the two interfering cw beams do not exit their interferometer coaxially and have to be realigned in a dedicated optical setup, which is itself stabilized using a second auxiliary laser.

Pump-probe setups in collinear geometry do not use metallic filters in the interferometer, since the annular generation and the inner dressing beams are recombined before the HHG gas jet \cite{14, Mairesse2003, 30weber}. In this case the cw laser can be directly superimposed to the main beams in the interferometer. However, in the end, this cw laser still has to be redirected to the detectors for the active stabilization, which requires at least one additional optic.

We can finally mention the active stabilization scheme proposed by Schlaepfer \textit{et al.} \cite{12}. The presence of an Optical Parametric Amplifier (OPA) in one of the arms avoids the need for an auxiliary laser. They instead collect the outer part of the IR generation beam passing around the filter and recombine it with the 1560 nm light exiting the OPA, in a BBO crystal optimized for second harmonic generation. The interference of the generation IR and the OPA light back-converted to 780 nm is used for the active stabilization. Yet, CEP stability of the two pulses is needed for this technique to work.

All  stabilization setups mentioned above require at least one additional optic, and often many more; hence the measured phase might fluctuate differently from the actual phase of interest between the IR and XUV beams. For instance, when the additional laser is guided on specific optics attached to the mounts of the main optics, the measured fluctuations do not include those caused by a thermal expansion of the main mirrors. Such residual power absorption of the femtosecond intense laser get all the more important as the repetition rate of the laser increases. In addition to that, existing active stabilization schemes increase the complexity of the optical setup.

Here we introduce a completely different approach, without any auxiliary beam or dedicated optical system, based on an error signal that directly depends on the XUV/IR delay in the active region of a gas phase electronic spectrometer. Specifically, we form an optical image of the point where the experiment is carried out, e.g. the sensitive region of a coincidence spectrometer. We collect and condition the two-photon ionization signal delivered by this auxiliary spectrometer, in order to obtain a measurement of the XUV/IR delay which is then used in a feedback loop. This technique offers an \textit{in situ} measurement and servo-locking of the delay of interest. In the following, we will refer to this  new technique as the LIZARD (Laser-dressed IoniZation for the Adjustment of the pump-pRobe Delay).
This article is organized as follows. In part \ref{sec:level2},  theoretical elements are introduced. Part \ref{experimental imp}  presents the experimental performances of the LIZARD, and the main conclusions and a few outlooks are discussed in part \ref{sec: conclusion}.

\section{\label{sec:level2}Theory}

\subsection{Pump-probe delay measurement}
\label{1.A}
Active stabilization techniques rely on interferometric signals that depend on the pump-probe delay. For instance, if we use a cw laser, a photodetector can be placed at the output of the interferometer and will deliver a signal which varies sinusoidally with the delay. A feedback loop can be performed so that this output signal is locked at zero \cite{5,8}. However, for applications requiring delay scans, it is of prime importance to be able to continuously lock the pump-probe phase at arbitrary set-points, with a uniform stability. 
A solution consists in using two signals modulated with the delay, but not in phase with one another \cite{1, 29, 31Freschi}. Calling $\phi(t)$ the phase difference between the pump and the probe beams, that we want to measure and stabilize, such signals write
\begin{equation}
\label{two signals}
  \begin{aligned}
  M_1(t) &= O_1 + A_1 \sin(\phi(t) + \Delta)  \\
  M_2(t) &= O_2 + A_2 \cos{\phi(t)}.
  \end{aligned}
\end{equation}
In the following, we assume that $M_1$ and $M_2$ are dimensionless quantities. A fast preliminary calibration scan over the delay range gives a measurement of the parameters $(O_1, O_2, A_1, A_2, \Delta)$, that we assume constant over the stabilization time. If $\Delta=0$, $M_1$ and $M_2$ are in phase quadrature; hence for any value of $\phi(t)$, either $M_1$ or $M_2$ will have a significant slope, ensuring good dynamics of the delay measurement. In the general case, measurements of $M_1$ and $M_2$ allow the reconstruction of the phase 
\begin{equation}
\label{arctan}
\phi(t)=\mathcal{U}\Bigg(\tan^{-1} \bigg(\frac{1}{\cos\Delta}\times\frac{A_2 \cdot (M_1(t) - O_1)}{A_1 \cdot (M_2(t) - O_2)}  -\tan\Delta\bigg) \Bigg),
\end{equation}
with $\mathcal{U}$ a phase unwrapping operator dealing with phase jumps, which works as long as the phase does not vary by more than $\pm \pi$ during the sampling time. More complex  solutions for the phase reconstruction exist, such as using more than two modulated signals \cite{3}.

Experimentally, due to detection noise, the values of $M_1$ and $M_2$ are known within some uncertainties $\delta M_1$ and $\delta M_2$. If $\Delta$ is not negligible, this leads to an uncertainty $\delta \phi$ on the phase measurement that strongly depends on the set-point phase. Indeed, $\delta \phi$ can be expressed as  (details in Appendix \ref{appendix})
\begin{eqnarray}
\label{delta phi}
\delta \phi=&&\sqrt{ \bigg( \frac{\partial \phi}{\partial M_1} \delta M_1\bigg)^2 + \bigg( \frac{\partial \phi}{\partial M_2} \delta M_2\bigg)^2}\nonumber\\
=&&\sqrt{ \bigg(\frac{\cos \phi}{A_1 \cos\Delta} \delta M_1 \bigg) ^2 +  \bigg(\frac{\sin(\phi+\Delta)}{A_2\cos\Delta} \delta M_2 \bigg)^2}.
\end{eqnarray}
\begin{figure}
\includegraphics[width = 9cm]{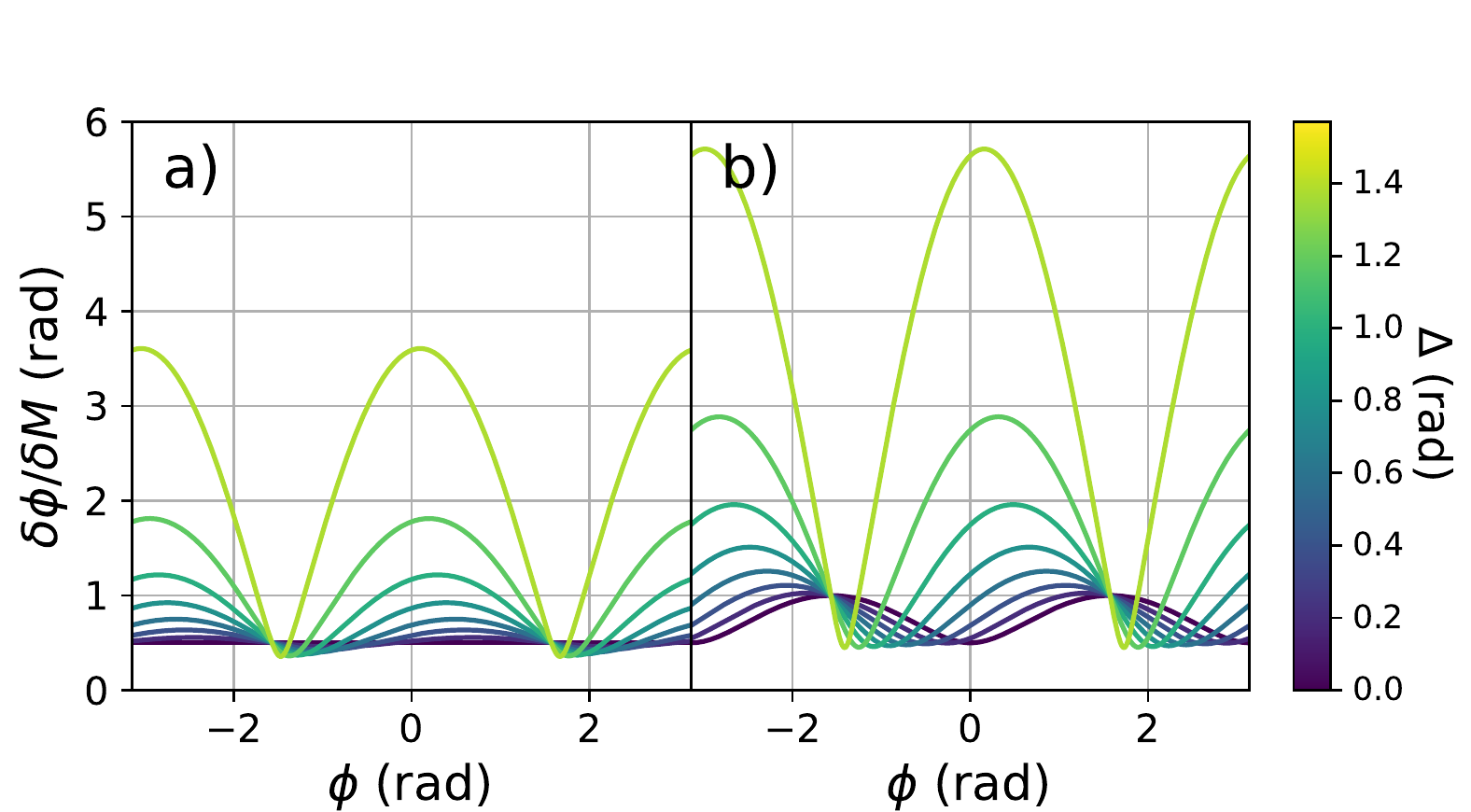}
\caption{(color online) Uncertainty on the reconstructed pump-probe phase due to the noise in the signals $M_1$ and $M_2$, for different values of $\Delta$. Both graphs were plotted for $\delta M_1 = \delta M_2 = \delta M$ and $A_1 = 2$. \textbf{a)} Case where $A_1 = A_2$ ($M_1$ and $M_2$ have the same signal to noise ratio). \textbf{b)} An example with $A_2 = A_1/2$ (the signal to noise ratio is lower for $M_2$ than for $M_1$).}
\label{fig_noise}
\end{figure}
The behavior of this function is shown in Fig.\,\ref{fig_noise} in some specific cases. If the two signals have equal signal to noise ratio (Fig.\,\ref{fig_noise}.a), $\delta \phi$ is uniform over the whole pump-probe phase range when $\Delta = 0$. This does not hold if $M_1$ and $M_2$ have different signal to noise ratios (Fig.\,\ref{fig_noise}.b). However, in both cases, the average value of $\delta \phi$ is minimized when $\Delta = 0$. The phase quadrature is therefore an optimal situation for a precise active stabilization. 

The originality of the LIZARD lies in the origins of the phase quadrature signals, which are extracted from the photoelectron spectrum obtained in a laser-dressed photoionization experiment.

\subsection{Principle of laser-dressed ionization by a HHG frequency comb}
\label{1.B}

\begin{figure}[h!]

\includegraphics[width = 5cm]{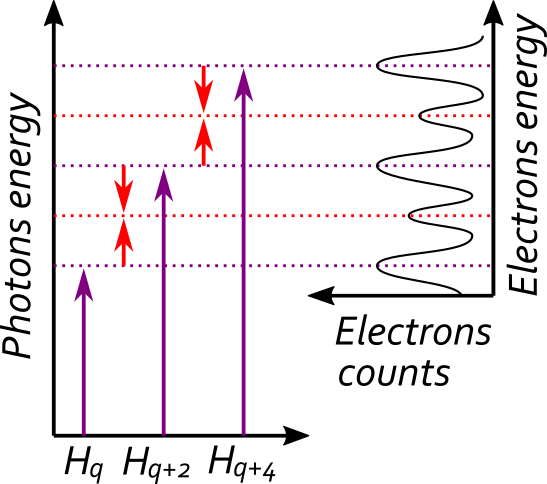}
\caption{(color online) Sketch of the two-photon transitions involved in laser-dressed ionization. XUV and IR photons are represented with purple and red arrows respectively. The photoelectron spectrum (right part) shows the main harmonics peaks, spaced by twice the energy of the driving laser. Weaker sidebands are located between those peaks, at even multiples of the driving laser energy.}
\label{fig_tpt}
\end{figure}

\begin{figure*}
\includegraphics[width =0.9\textwidth]{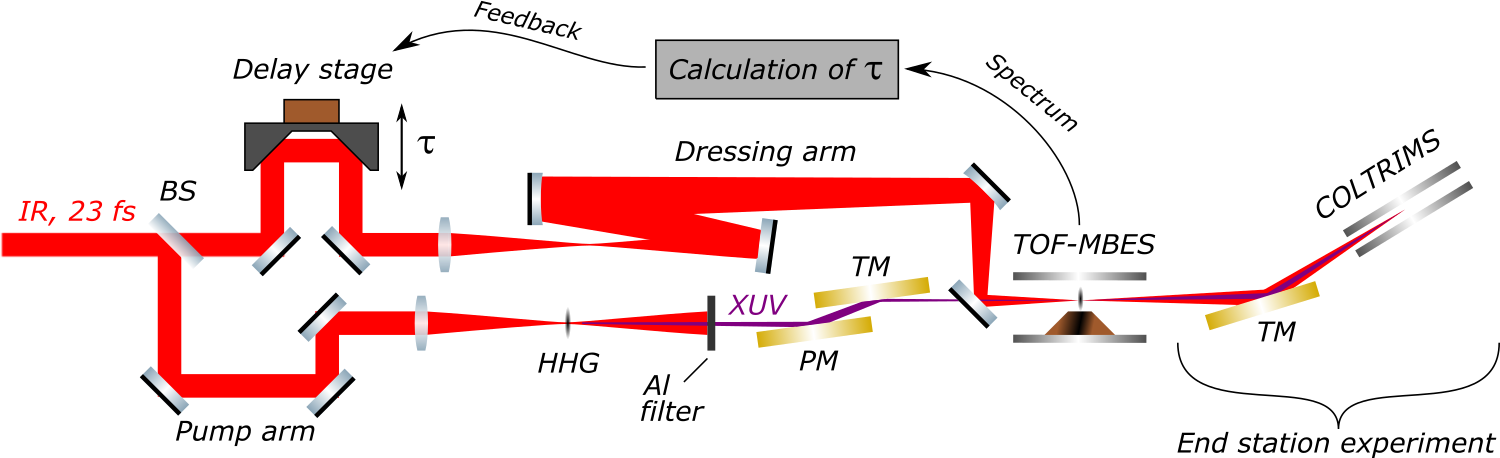}
\caption{\label{fig:wide}(color online) Experimental setup. BS: beam splitter, PM: plane mirror, TM: toroidal mirror. The interferometer arms are approximately 2.5 m long. The Cold Target Recoil Ion Momentum Spectrometer (COLTRIMS) experiment in the end station is a possible application of the LIZARD; it is not part of the stabilization system itself.}
\label{fig_setup}
\end{figure*}

The setup is similar to generic pump-probe experiments in attosecond physics (see Fig.\,\ref{fig_setup}). The IR laser (of angular frequency $\omega_L$) is separated into two beams using a beam splitter. Most of the intensity is used for HHG, but a small portion, called "dressing", propagates unchanged. The two beams are then recombined and focused in a Time Of Flight Magnetic Bottle Electronic Spectrometer (TOF-MBES). With multicycle driving laser pulses, the photoelectron energy spectrum exhibits peaks located at the harmonics energies.
The field in the gas jet being a superposition of XUV and IR, two-photon transitions are also allowed. They are represented for different harmonics in Fig.\,\ref{fig_tpt}. Two transitions -- absorption of a photon from harmonic $H_q$ and an IR photon; absorption of a photon from harmonic $H_{q+2}$ and stimulated emission of an IR photon -- yield the same photoelectron energy (red dashed lines in Fig.\,\ref{fig_tpt}), creating sidebands between the harmonic peaks in the photoelectron spectrum.

These two quantum paths interfere, which leads to a modulation of the sidebands intensity with the time delay $\tau$ between the XUV and IR fields, at twice the laser frequency \cite{13}
\begin{equation}
\label{equation sb}
    I_q(\tau) = O_q + A_q\sin{(2\omega_L \tau +\phi_\text{at} + \phi_{q+1} - \phi_{q-1})},
\end{equation}
with $I_q$ the intensity of sideband $SB_q$ (at energy $q\omega_L$), $\phi_\text{at}$ an usually negligible intrinsic atomic phase, and $\phi_q$ the spectral phase of the XUV field at energy $q\omega_L$. The term $\phi_{q+1} - \phi_{q-1}$ in Eq.\,\eqref{equation sb} is therefore the increment of the XUV spectral phase between the two successive harmonics surrounding sideband $SB_q$. $O_q$ and $A_q$ are the offset and amplitude of $SB_q$, which depend on the intensity of harmonics $H_{q-1}$ and $H_{q+1}$, the intensity of the dressing beam, and detection settings.

\begin{figure}
\includegraphics[width = 9cm]{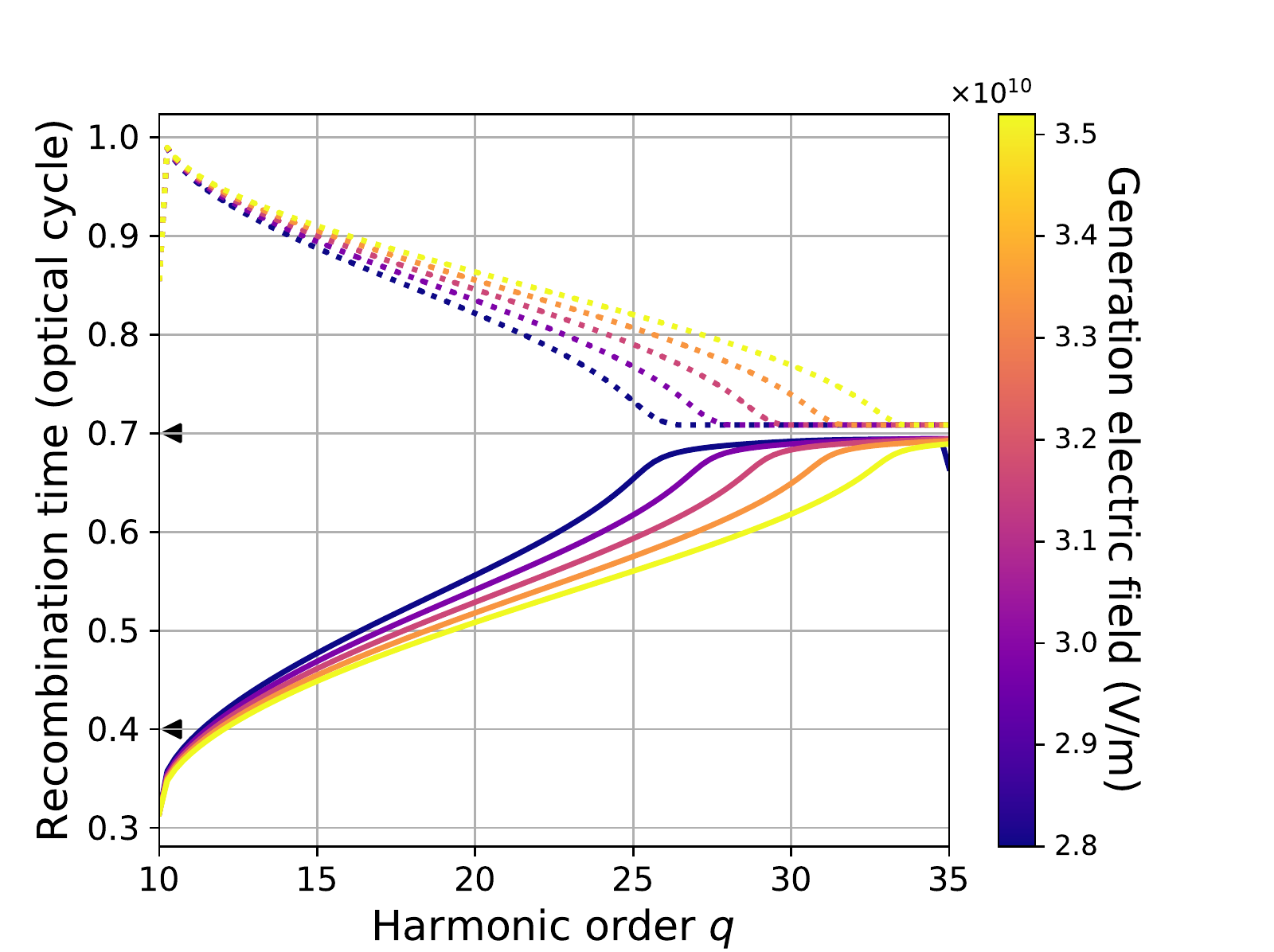}
\caption{(color online) Electron recombination time calculated within the strong field approximation \cite{32Lewenstein}, as a function of the harmonic order. We considered an Argon atom submitted to a 800 nm wavelength. We disregarded any envelope effect of the driving pulse. The short and long trajectories are represented with continuous and dotted lines respectively. The simulation was performed for different values of the IR laser electric field. The two triangles on the vertical axis indicate roughly the range of the XUV emission time in the case of short trajectories.}
\label{fig_energy} 
\end{figure}

The modulation of the sidebands with the pump-probe delay $\tau$, coupled to the fact that they are not in phase with one another, make them excellent candidates for an active stabilization of the delay. Quite remarkably, the physics of HHG, which gives rise to an 'attosecond chirp'\cite{16}, will often provide sidebands sufficiently separated in phase so that at least two of them  can be found close to phase quadrature. Indeed, the three steps model of HHG -- in which an outer electron successively undergoes tunnel ionization, acceleration in the continuum and recombination with the parent ion --  predicts that the recombination time of the electron will vary almost linearly with the photon energy of the emitted harmonic (Fig.\,\ref{fig_energy}). The recombination time difference between the highest and the lowest part of the emitted spectrum (represented with two triangles in Fig.\,\ref{fig_energy}) is approximately $0.3\,T_L$, with $T_L$ the IR optical cycle, and does not depend on the generating laser intensity. The corresponding dephasing between the highest and the lowest sidebands is thus given by $\Delta \Phi \approx 2\omega_L \times 0.3\,T_L = 1.2\,\pi$. As it is greater than $\pi/2$, whatever the generating medium and the laser intensity, one should always be able to find two sidebands in the photoelectron spectrum whose phase difference is close enough to $\pi/2$ to serve as the $M_1$ and $M_2$ signals in the LIZARD scheme. As a downside, this also means that attosecond pulses which are perfectly post-compressed (by dedicated devices) would not allow a straightforward implementation of the LIZARD, since they do not possess chirp.

It may be noted that the high harmonic peaks can be used as well: they are also modulated with the delay, with a phase just opposite to that of the neighbouring sidebands. This is why, in the following, 'delay modulated signals' will indifferently refer to sidebands or harmonics.

\section{\label{experimental imp}Experimental implementation and results}

To test the LIZARD scheme, experiments were conducted on the user-oriented FAB10 beamline at the ATTOLab facility. The beamline is based on a Ti:Sapphire chirped pulse amplification laser \cite{Golinelli} delivering 23 fs pulses carrying 2\,mJ, with a 10 kHz repetition rate. Focused onto a continuous Argon gas jet, it drives a HHG source emitting attosecond pulses trains (APT).  The beamline is optimized for experiments requiring fairly high recurrence and long acquisition times, for instance coincidence measurement in gas phase \cite{JJ} or photoemission in solids \cite{21}. 
RABBIT \cite{14,15} spectroscopy is permanently operational on FAB10, which allows the characterization of the XUV light and the experiments in the end station to be performed simultaneously. For that purpose, the XUV beam is first focused in the TOF-MBES, then re-focused by a gold mirror in the user’s experimental chamber. The parity of the number of mirrors and focusing optics is identical in the two arms of the interferometer

To make sure that we are sensitive to delay fluctuations only, we have to reduce as much as possible the noise on the signals $M_1$ and $M_2$. It may be due to fluctuations of the dressing intensity and the XUV harmonic intensity. The latter will cause changes in the total number of emitted photoelectrons per second. To limit this effect, a beam pointing stabilization of the IR the laser is performed. Moreover, the TOF-MBES signal will be averaged over a large number of laser shots (which is possible thanks to the high repetition rate of the beamline), and normalized. The resulting value at energy $q\omega_L$ is denoted $\hat{I}_q$.

The time delay $\tau$ between the IR dressing and the XUV probe is controlled by two mirrors mounted on a translation stage (model SmarAct SLC-24 150-P-R-S, based on a stick-slip technology combining long travel range (10 cm) and piezo accuracy (1 nm)) before the HHG chamber (Fig.\,\ref{fig_setup}).
The electronic signal from the magnetic bottle's  microchannel plate (MCP) detector is acquired with an oscilloscope (model Lecroy Waverunner 6 Zi).

Before starting the stabilization, a fast delay scan is performed in order to record the oscillations of the two selected sidebands. The parameters $O_1$, $O_2$, $A_1$, $A_2$ and $\Delta$ of Eq.\,\eqref{two signals} are extracted from sinusoidal fits of the data. During the stabilization process, the feedback loop performed at each step consists in reconstructing the pump-probe phase difference $\phi(t)$ using Eq.\,\eqref{arctan}, calculating the error value  $\Delta\phi(t) = \phi(t) - \phi_0$ (with $\phi_0$ the desired set-point phase) and performing feedback using a PID (proportional-integral-derivative) filter. Those operations are performed by a GUI program based on PyMoDAQ \cite{pymodaq}, an open source Python framework for data acquisition programs. Since we mostly need to compensate for slow thermal delay drifts, a 10 Hz correction frequency is high enough. It allows us to average the photoelectron signal over 1000 laser shots, ensuring a reasonable level of noise. 

\subsection{Stabilization at a fixed set-point}

\begin{figure*}

\includegraphics[width =0.95\textwidth]{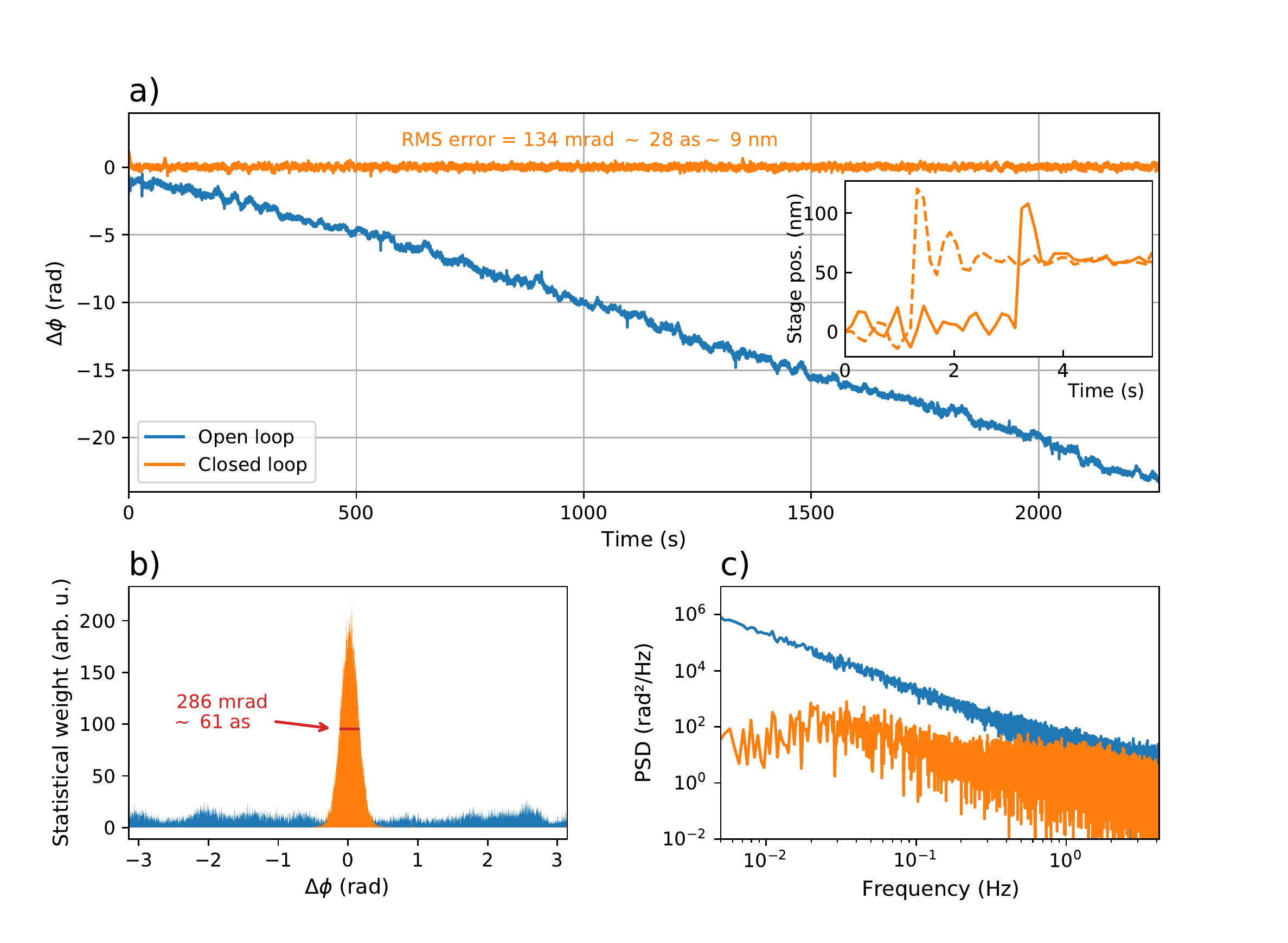}
\caption{(color online) Comparison between open loop (blue) and closed loop (orange) operation. \textbf{a)} Long term measurement of the phase error $\Delta\phi(t) = \phi(t) - \phi_0$. The inset shows the evolution of the delay stage position when the setpoint optical paths difference is changed by 100 nm, in the case of a non-optimized (dashed line) and optimized (continuous line) PID. As expected, the kick yields an overall 50 nm displacement of the stage (because of the mirrors corner cube configuration). For an optimized PID, we consider that the new setpoint value is reached in approximately 0.5 s. \textbf{b)} Phase error histogram. The full width at half maximum for the closed loop error is indicated with a red line. Assuming a gaussian distribution, it should be approximately $2\sqrt{\ln 2} \approx 1.67$ times the RMS error (224 mrad). \textbf{c)} Power spectral density of the phase error $\delta \phi(t)$. The low frequency noise, corresponding to long term drifts, is efficiently suppressed by the LIZARD.}
\label{fig_results}
\end{figure*}

A typical long-term LIZARD measurement of the pump-probe phase difference, in open loop, is shown in Fig.\,\ref{fig_results}.a (blue curve). We observe a drift which is remarkably linear, with a rate of
\unit{36}{\rad \per \hour} $\sim$ \unit{7.6}{\femto \second \per \hour} $\sim$ \unit{2.3}{\micro \meter \per \hour}, that we attribute to the thermal expansion of the mirrors. In other words, the difference of optical paths varies by a sideband spatial period in 10 minutes. This passive stability is too poor to maintain attosecond precision during experiments lasting several hours, forcing the use of convoluted post-analysis protocols \cite{JJ}.

The measurement in closed loop is shown as the orange curve in Fig.\,\ref{fig_results}.a. The two modulated signals chosen for the active stabilization are $SB_{14}$ and $SB_{22}$. They are separated in phase by $0.46\, \pi \,$ rad, very close to phase quadrature. We reached a phase stability of 134 mrad RMS, equivalent to a 28 as time delay stability and a 9 nm optical path stability, which is competing with state of the art performance \cite{30weber, 12, 8, 2, 3, 4, 6}. Our active stabilization system turns out to be robust and easy to operate. After a proper alignment of the beamline, it only takes one minute to select two energy bands, make a calibration scan and launch a long-term stabilization.

\begin{figure*}

\includegraphics[width =\textwidth]{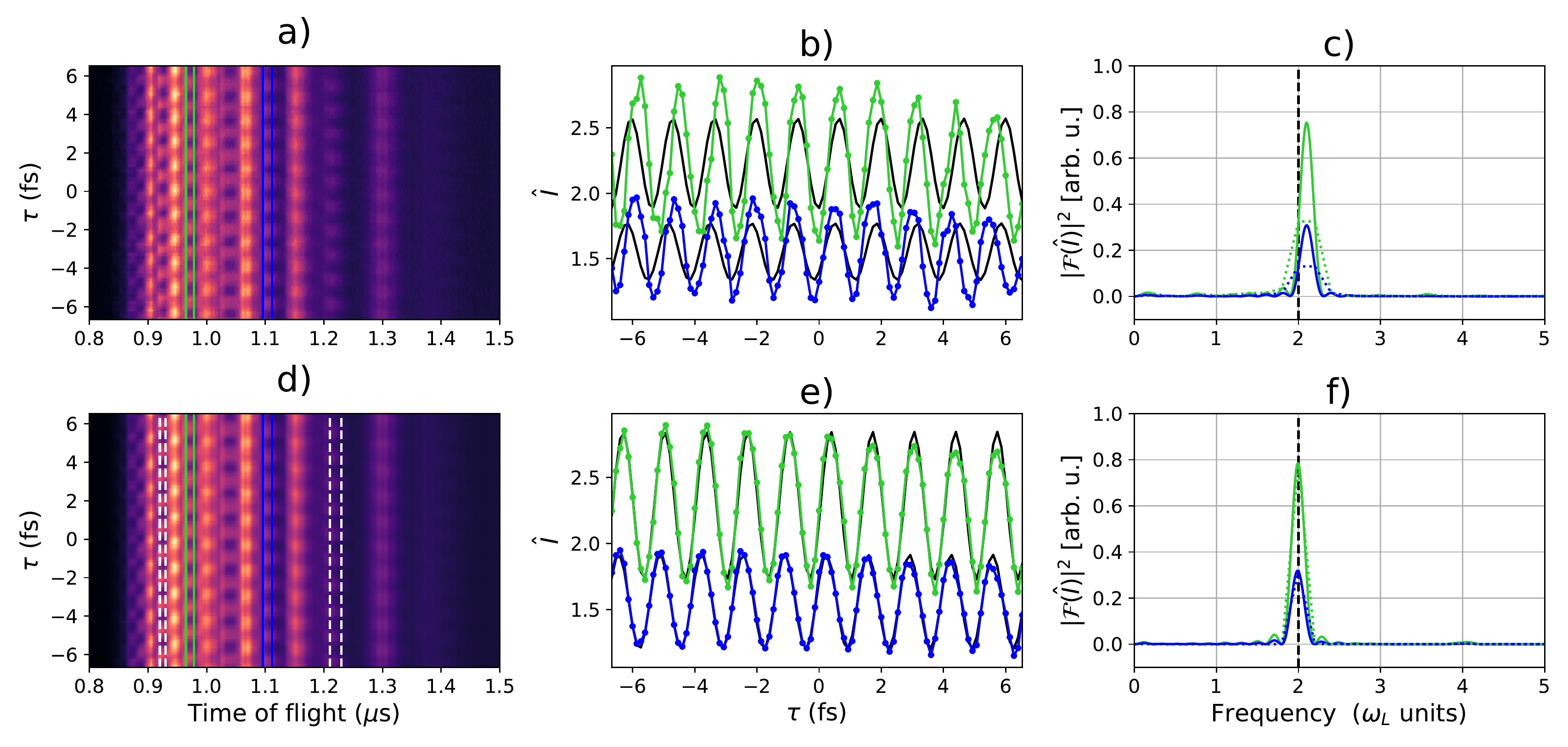}
\caption{(color online) Comparison between a classical open loop RABBIT scan  (\textbf{a},\textbf{b},\textbf{c}) and a RABBIT scan stabilized by LIZARD (\textbf{d},\textbf{e},\textbf{f}). (\textbf{a},\textbf{d}) Modulation of the photoelectron spectrum with the delay $\tau$, after background subtraction and normalization. In the stabilized scan (d), $\hat{I}_\text{14}$ and $\hat{I}_\text{22}$ (white dotted bands) were used as the two modulated signals $M_1$ and $M_2$ for the active stabilization. (\textbf{b},\textbf{e}) $\hat{I}_\text{16}$ (blue) and $\hat{I}_\text{20}$ (green), as a function of the delay. The signals are integrated over the width of the bands with the corresponding colors shown in a) and d). The black curves are sinusoidal fits. (\textbf{c},\textbf{f}) Squared modulus of the Fourier transforms of $\hat{I}_\text{16} (\tau)$ and $\hat{I}_\text{20}(\tau)$. The average value of each of the two modulated signal was first subtracted in order to remove very low frequency components. The vertical dashed line shows the theoretical $2\omega_L$ sidebands angular frequency. The raw Fourier transforms (dotted lines) were smoothed using zero padding (solid lines).}

\label{fig_scans}
\end{figure*}

\subsection{\label{stabilized scans}Stabilized pump-probe delay scans}

Fig.\,\ref{fig_results} presented the case of an interferometer locked at a constant value of the pump-probe phase. The phase can however be locked at any arbitrary desired set-point. In order to demonstrate the capabilities of the technique, we performed open loop and closed loop RABBIT scans. In a classical open loop RABBIT scan, the pump-probe optical path difference is typically scanned over $4\, \micro \meter$ with $40\,\nano \meter$ steps (Fig.\,\ref{fig_scans}.a). Such a scan ideally spans over exactly 10 sideband periods, but is distorted by delay drifts (Fig.\,\ref{fig_scans}.b). As the delay drift  in our experiment is primarily linear, it results in a shift of the frequency of the sidebands oscillation (Fig.\,\ref{fig_scans}.c). Using a least squared method, we fitted the data with the sinusoidal function defined by Eq.\,\eqref{equation sb}. Importantly, the sidebands frequency is not a parameter of the fit which adjusts only the offset, amplitude and phase.

\begin{figure}

\includegraphics[width = 8.7cm]{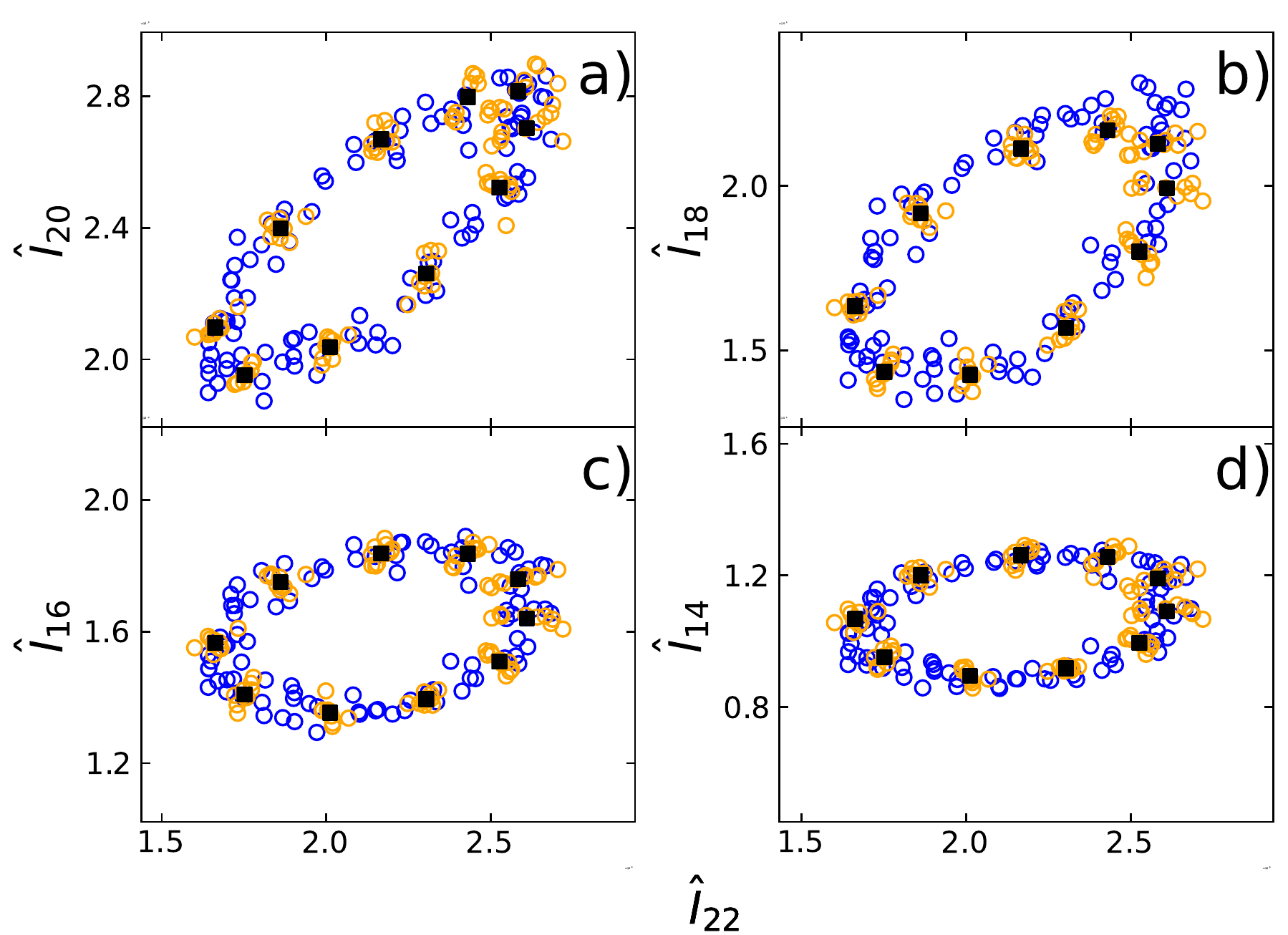}
\caption{(color online) Intensity correlation between four different sidebands extracted from the spectrum of Fig.\,\ref{fig_scans}, and sideband $SB_{22}$. Data corresponding to the open loop scan and to the closed loop scan are in blue and orange respectively. Black squares indicate the average positions of the sets of data corresponding to the same set-point phase. All axis have equal unit lengths. The increasing tilt of the ellipses is due to the attochirp. \textbf{a)} $SB_{20}$ versus $SB_{22}$. \textbf{b)} $SB_{18}$ versus $SB_{22}$. \textbf{c)} $SB_{16}$ versus $SB_{22}$. \textbf{d)} $SB_{14}$ versus $SB_{22}$. Those two particular sidebands were used for the LIZARD stabilization, they are in phase quadrature and yield a non-tilted correlation ellipse.}
\label{fig_ellipse}
\end{figure}

Fig.\,\ref{fig_scans}.d is a LIZARD-stabilized RABBIT scan, obtained with the following procedure. The optical paths difference is stabilized and the set-point is regularly adjusted by steps of 40 nm. The optimization of the PID ensures that the new set-point value is reached in only a few tenth of seconds. Data corresponding to the same set-point are averaged in post-analysis. In the experiment presented here, the phase remained locked at each set-point during 50 photoelectron spectrum acquisitions. For the relevancy of the comparison, both the non stabilized and the stabilized scan had the same duration, approximately 5 minutes. With the active stabilization, the delay scan matches the sinusoidal fit almost perfectly (Fig.\,\ref{fig_scans}.e). The sidebands Fourier transform (Fig.\,\ref{fig_scans}.f) contains a thin peak nicely centered at $2\omega_L$. The stabilized scan approaches the perfect sinusoidal oscillations of Eq.\,\eqref{equation sb}. Another way to visualize the effect of the LIZARD stabilization is to look at the correlation between two sidebands intensities (Fig.\,\ref{fig_ellipse}), similarly to diagnosis performed with FEL \cite{26} and the recently developed TURTLE \cite{25}. Since the pump-probe delay is now precisely adjusted, the stabilization results in a squeezing of the data points around the phase set-points.

\begin{figure*}

\includegraphics[width =1.03\textwidth]{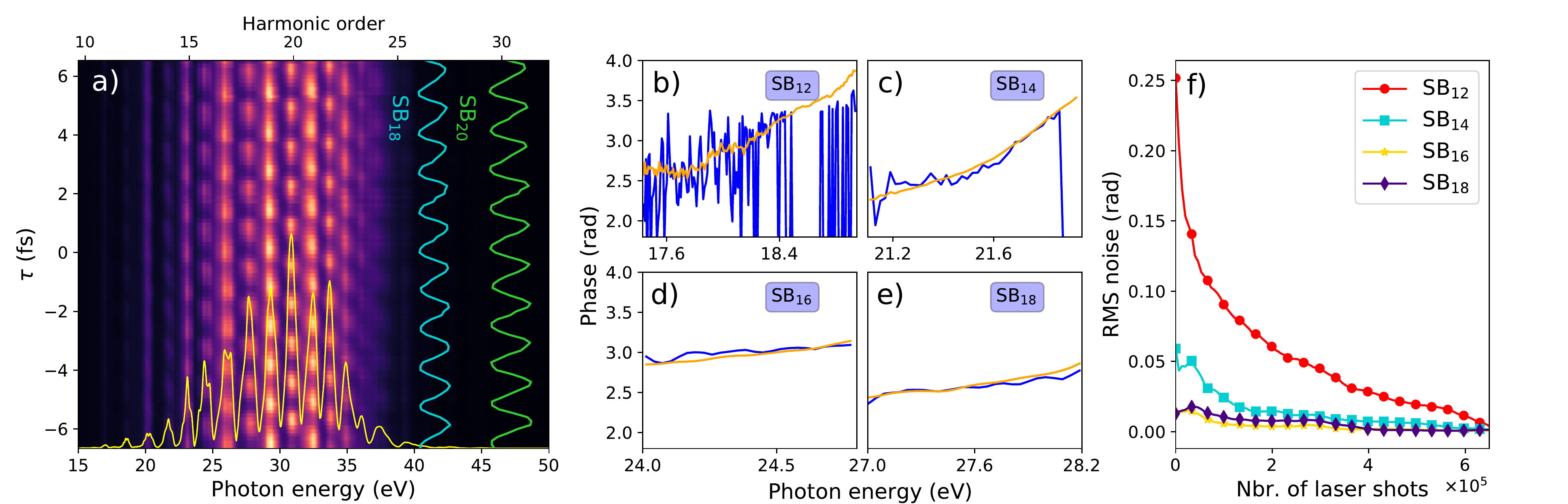}
\caption{(color online) \textbf{a)} RABBIT scan in argon stabilized by LIZARD that lasted approximately 2 hours. Between each 20 nm optical paths change, spectra were accumulated during  65 s, which corresponds to $6.5 \times 10^5$ laser shots. The yellow curve shows the $2 \omega$ amplitude. The very sinusoidal oscillations of two of the sidebands (SB$_{18}$ and SB$_{20}$) are plot as an example in blue and green.  \textbf{(b-c-d-e)} Spectral resolution of some sidebands phases (Rainbow RABBIT technique \cite{24}), in the case of the 2 hours long stabilized scan (orange) and a fast (30 s) calibration scan recorded just before (blue). \textbf{f)} RMS noise of the spectral phase measurement, with respect to the number of laser shots over which the spectra were averaged during each scan step.} 
\label{fig_longscan}
\end{figure*}


One could argue that we should not be surprised to observe almost perfect sinusoidal sidebands with 2$\omega_L$ angular frequency in the case of closed loop RABBIT scans, since we used the RABBIT photoelectron spectrum itself to perform the stabilization. An important point is that the active stabilization is not using the whole photoelectron signal but only two specific energy bands. $SB_{16}$ and $SB_{20}$, studied as an example in Fig.\,\ref{fig_scans}.d-f, are not the sidebands used for the LIZARD as the input signals $M_1$ and $M_2$. We can therefore consider that the information carried by the remaining sidebands is preserved in a stabilized scan. The comparison between Fig.\,\ref{fig_scans}.b and Fig.\,\ref{fig_scans}.e confirms that the sidebands sinusoidal fit, calculated with an imposed angular frequency of $2\omega_L$, matches the experimental data more precisely in the case of a stabilized RABBIT scan. Thus the estimation of the phase difference $\Delta \Phi_\text{16-20}$ between $SB_{16}$ and $SB_{20}$ (which corresponds to a photoemission time delay between two harmonics) will be more accurate in the case of a stabilized scan. To verify this, we use our sinusoidal fit to estimate the uncertainty on the measurement of $\Delta \Phi_\text{16-20}$. The standard deviation associated to the measured phase of a given sideband is obtained by considering the variance of the sideband data points equal to the variance of the residual (the distance between the data and the fit). Let us call $\Delta \Phi_\text{16-20}^\text{o-l}$ and $\Delta \Phi_\text{16-20}^\text{c-l}$ the phase difference between the two sidebands extracted from the open loop and the closed loop RABBIT scan respectively. We found, with a $2\sigma$ uncertainty, $\Delta \Phi_\text{16-20}^\text{o-l} = \unit{0.64}{} (\pm \unit{0.55})\rad$ and $\Delta \Phi_\text{16-20}^\text{c-l} = \unit{0.67} (\pm \unit{0.06})\rad$. Here, the active stabilization reduces the uncertainty on the measurement by a factor of 8. 

Finally, Fig.\,\ref{fig_longscan} illustrates the feasibility of delay scans lasting several hours. It is readily visible from Fig.\,\ref{fig_longscan}.b-e that accumulating long spectra while stabilizing with LIZARD drastically increases the precision of spectral phase retrieval, especially at low energies, where the $2 \omega$ amplitude is weak. To quantify this, we computed the RMS distance between the spectral phase resulting from the averaging over intermediate accumulation times, and the spectral phase obtained by averaging over 65 s, considered as a reference (orange curves in Fig.\,\ref{fig_longscan}.b-e). The results are displayed in Fig.\,\ref{fig_longscan}.f. As expected, although increasing the accumulation time has a negligible effect where the signal to noise ratio is high ($SB_{16}$ and $SB_{18}$), the benefit is dramatic for the weak signals of $SB_{12}$ and $SB_{14}$. 
Because it allows such long averaging times, we can expect the LIZARD to become a precious tool for the measurement of spectral phases when the signal to noise ratio is particularly poor, due for instance to weak cross sections of the channels of interest \cite{Klunder2011}. 


\section{Conclusion and outlook}
\label{sec: conclusion}
Our technique is particularly suited to attosecond beamlines that are already equipped with an electron time of flight spectrometer. It is fundamentally different from existing active stabilization schemes, which use the interference of photons, and measure a pump-probe phase difference accumulated during the propagation of a cw laser in the interferometer. The LIZARD technique does not require any additional cw laser or even additional optical elements but a refocusing optic that is common to both beams. Its principle relies on the interference of photoelectron wavepackets, therefore providing an \textit{in situ} measurement of the XUV/IR phase difference in the active region of the electronic spectrometer. The underlying physics of HHG guarantees that the spectrum contains two sidebands in phase quadrature in the vast majority of generating conditions. This allowed us to stabilize the delay with a 28 as RMS accuracy during several hours - which is comparable to state of the art stability performance - and to adjust the set-point delay over a few microns range, thus enabling pump-probe experiments requiring long acquisition times. We also showed that this technique can bring higher precision in the measurement of photoemission time delays by the RABBIT technique. 
From a practical perspective, the simplification of the optical system is counterbalanced by the need of an electron spectrometer. This device is however quite common on attosecond beamlines, and could even be used in a simpler version collecting only two time of flight windows, one for each modulated signal. 

Further improvements of the method are intended. We could use the whole photoelectron spectrum instead of only two sidebands, so that the pump-probe delay could be deduced from a least squared optimization of all the sidebands intensity. A statistical study showed that the accuracy of the delay measurement increases with the number of sidebands considered.
We could also increase the correction frequency by using analog electronics instead of a numerical code. However, since this would reduce the number of averaged shots per delay measurements, it is not certain that it would improve the stability. 
This novel method could possibly be extended to active stabilization schemes with isolated attosecond pulses (IAP). Although they exhibit a continuous frequency spectrum, the standard IAP temporal reconstruction techniques, FROG-CRAB \cite{23} and PROOF \cite{28}, are also based on laser-dressed photoionization, and yield delay-modulated spectrograms. By integrating the spectrum over well chosen energy bands, one might extract the required phase quadrature delay-modulated signals, and use the LIZARD to actively stabilize pump-probe experiments with IAP.


Finally, we believe that this approach could be applied to the stabilization of two-color interferometers in other spectral domains, as long as one of the beams is able to photoionize a target. Here we described in details the situation where one of the two pulses is in the XUV spectral range, therefore photoionization is possible thanks to the high energy of the photons involved. However, even at lower photons energies (for instance, consider an infrared pulse and its second harmonic), photoelectrons and oscillating sidebands could be obtained by above-threshold ionization of gas targets \cite{Agostini, BeaulieuATI}.

\begin{acknowledgments}
The authors acknowledge Bertrand Carré,  Carlo Spezzani, François Polack, David Dennetiere, Ismaël Vadillo-Torre, Michel Bougeard and Julien Lenfant who contributed to the design and construction of the FAB10 beamline; and Pascal D'Oliveira and Fabien Lepetit for the design, implementation and operation of the laser.

The laser system and the experimental setup are supported by the French “Investments for the Future” of the Agence Nationale pour la Recherche (Contract No. 11-EQPX0005-ATTOLAB and ANR-14-CE32-0010 - Xstase), the Scientific Cooperation Foundation of Paris-Saclay University through the funding of the OPT2X research project (Lidex 2014), by the Île-de-France region through
the Pulse-X project and by the European Union’s Horizon 2020 Research and Innovation Programme No. EU-H2020-LASERLAB-EUROPE-654148.

\end{acknowledgments}

\appendix

\section{Derivation of Eq.\,\eqref{delta phi}}
\label{appendix}

The tangent of the reconstructed phase $\phi$ writes, according to Eq.\,\eqref{two signals}

\begin{equation}
    \tan \phi = \frac{1}{\cos\Delta}\times \bigg(\frac{M_1 - O_1}{M_2 - O_2} \times \frac{A_2}{A_1}\bigg)  -\tan\Delta
\end{equation}
Let us differentiate this equation with respect to $M_1$

\begin{equation}
\frac{\partial \tan \phi}{\partial M_1} = \frac{1}{\cos\Delta}\bigg(\frac{A_2}{A_1(M_2 - O_2)} \bigg)  = \frac{1}{A_1 \cos\phi \cos\Delta}
\end{equation}
Similarly, we obtain for the derivative with respect to $M_2$

\begin{equation}
\frac{\partial \tan \phi}{\partial M_2} = \frac{-\sin(\phi+\Delta)}{A_2 \cos^2\phi \cos\Delta}
\end{equation}

The derivatives of $\phi$ with respect to $M_1$ and $M_2$ are obtained using $ \text{d}\phi = \cos^2\phi \times \text{d}(\tan \phi)$, yielding the following equations

\begin{equation}
\label{two expr}
  \begin{aligned}
  \frac{\partial \phi}{\partial M_1} &= \frac{\cos \phi}{A_1 \cos\Delta}  \\
  \frac{\partial \phi}{\partial M_2} &= - \frac{\sin(\phi+\Delta)}{A_2 \cos\Delta}
  \end{aligned}
\end{equation}

By definition, the uncertainty on the phase measurement is

\begin{equation}
\label{def delta phi}
\delta \phi = \sqrt{ \bigg( \frac{\partial \phi}{\partial M_1}  \delta M_1\bigg)^2 + \bigg( \frac{\partial \phi}{\partial M_2}  \delta M_2\bigg)^2}
\end{equation}

We obtain the last term of Eq.\,\eqref{delta phi} by injecting Eq.\,\eqref{two expr} in Eq.\,\eqref{def delta phi}.


\bibliographystyle{apsrev4-1}

\end{document}